\documentclass[twoside,12pt]{article}

\usepackage{amssymb,endnotes}

\newcounter{muni}
\newenvironment{remunerate}{\begin{list}{{\rm \arabic{muni}.}}
{\usecounter{muni}\setlength{\leftmargin}{0pt}
\setlength{\itemindent}{38pt}}}{\end{list}}

\newcommand{\nc}{\newcommand}                  \nc{\nf}{\infty}    
\nc{\dst}{\displaystyle}                       \nc{\nnb}{\nonumber} 
\nc{\beq}{\begin{equation}}                    \nc{\eeq}{\end{equation}} 
\nc{\beqa}{\begin{eqnarray}}                   \nc{\eeqa}{\end{eqnarray}}
\nc{\brm}{\begin{remunerate}}                  \nc{\erm}{\end{remunerate}}
\nc{\barr}{\begin{array}}                      \nc{\earr}{\end{array}}
\newtheorem{theorem}{Theorem}             \nc{\mc}{\mathcal}

\nc{\bs}{\backslash}        \nc{\nl}{\newline}      \nc{\mb}{\mathbb}
\nc{\qq}{\quad\quad}        \nc{\ol}{\overline}     \nc{\pt}{\partial}
\nc{\dg}{\dagger}

\nc{\alf}{\alpha}    \nc{\be}{\beta}        \nc{\ga}{\gamma}    
\nc{\de}{\delta}     \nc{\eps}{\epsilon}    \nc{\vtht}{\vartheta}
\nc{\om}{\omega}     \nc{\vp}{\varphi}    \nc{\vsi}{\varsigma}
\nc{\vrho}{\varrho}  \nc{\tht}{\theta}      \nc{\la}{\lambda}

\nc{\Om}{\Omega}     \nc{\Ga}{\Gamma}      \nc{\De}{\Delta}

\nc{\Log}{{\rm Log }}           \nc{\tg}{{\rm tg }}
\nc{\sh}{{\rm sh }}             \nc{\ch}{{\rm ch }}

\nc{\tr}{{\rm tr }}

\nc{\LIM}{\mathop{\smash{\rm LIM}}}

\title{\bf Self-adjoint extensions of operators and the teaching of quantum 
mechanics}
\author{\\Guy BONNEAU
\thanks{\noindent Laboratoire de Physique Th\'eorique et des Hautes Energies,
 Unit\'e associ\'ee au CNRS UMR 7589,~Universit\'e Paris 7-Denis Diderot,
 2 Place Jussieu, 75251 Paris Cedex 05.}
$\hspace{1.5cm}$ Jacques FARAUT
\thanks{\noindent Laboratoire d'Analyse Alg\'ebrique,~Universit\'e  Paris
6-Pierre et Marie Curie,
 4 Place Jussieu, 75252 Paris Cedex 05.}
\\ \\ Galliano VALENT$^{~*}$\\ \\}
\date{ }
\begin{document}
\maketitle

\begin{abstract}
\noindent For the example of the infinitely deep well potential, we point out 
some paradoxes which are solved by a careful analysis of what is a truly self-
adjoint operator. We then describe the self-adjoint extensions and their 
spectra for the momentum and the Hamiltonian operators in different physical 
situations. Some consequences are worked out, which could lead to  experimental
checks.
\end{abstract} 

\vfill {\bf PAR/LPTHE/99-43}\hfill  December 1999
\newpage
\section{Introduction} In most French universities, quantum mechanics is
usually taught in the  third year courses, separately from its applications to
atomic, molecular and  subnuclear  physics, which are dealt with during the
fourth year. In such  ``first contact" lectures, many mathematical subtleties
are necessarily left  aside. However, even in such commonly used examples as
infinitely deep  potential wells, overlooking the mathematical problems leads to
contradictions  which may be detected by a careful student and which have to do
with a precise  definition of the ``observables" i.e. the self-adjoint operators.

Of course, experts in the mathematical theory of unbounded operators in Hilbert 
spaces know the correct answer to these questions, but we think it could be 
useful to popularize these concepts among the teaching community and the more 
mature students of fourth year courses. In particular, the role of the boundary 
conditions that lead to self-adjoint operators is missed in most of the
available textbooks, the one by Ballentine \cite[p. 11]{Ba} being a notable
exception as  it includes a discussion of the momentum operator. But there, we find only two references relevant to the 
subject. The first one \cite{Ca} considers a particular self-adjoint 
extension of the momentum and of the Hamiltonian for a particle in a 
box, which is interpreted as describing a situation with spontaneous 
symmetry breaking. The second one \cite{CLM} mentions the self-adjoint 
extensions of the Hamiltonian for a particle in a semi-axis and its 
relevance, first pointed out by Jackiw \cite{Ja}, to the renormalization 
of the two dimensional delta potential.

The aim of this paper is to emphasize the importance of the boundary conditions 
in the proper definition of an operator and to make available to an audience of 
physicists basic results which are not so easily extracted from the  large amount of mathematical literature on the subject.

The paper is organised as follows : in Section 2 we discuss some paradoxes  met
in the study of the infinite potential well. Then, in Section 3, we present  a
first analysis of the boundary conditions for the self-adjoint extensions  of
the momentum operator.

In Section 4 we introduce the concept of deficiency indices and state von 
Neumann's theorem. In Section 5 we apply it to the self-adjoint extensions of
the  momentum operator for which the spectra, the eigenfunctions and some
physical  consequences of these are given. We hope that, despite some
technicalities  needed for precision (which can be omitted in a first reading),
the  results are of easy access. The reader interested in these
technical  aspects may consult the references \cite{AG} and \cite{Na}.

Then, in section 6, we describe the self-adjoint extensions of the Hamiltonian 
operator in various settings (on the real axis, on the positive semi-axis and 
in a box). Several physical implications are analysed, while in section 7, we 
use different constraints from physics to reduce the set of all possible 
self-adjoint extensions.

We have gathered in appendix A some technical details on the extensions of
the momentum operator and in appendix B we discuss the spectra of the 
Hamiltonian operator for a particle in a box. A proof for parity preserving self-adjoint extension is given in Appendix C.

\section{The infinite potential well : paradoxes} Let us consider the standard
problem (see for example  \cite[p. 299]{JMLL} or  
\cite[p. 109]{Gr}) of a particule of mass  $m$ in a one dimensional, infinitely 
deep, potential well of  width $L$ :
\beq\label{a1} V(x) = 0\ ,\ \ x\in ]-\frac{L}{2},\ +\frac{L}{2}[\ \ ;
\ \  \ \ \ V(x)=\ \infty\ ,\ \ |x|\ge \frac{L}{2}\ .
\eeq Stationary states are obtained through the Schr\" odinger (eigenvalue)
equation 
$$H\phi(x) = E\phi(x)$$ 
\underline{and the vanishing of their wave function at both ends}. This means 
that the action of the ha\-miltonian operator for a free particle, unbounded  on
the closed interval $[-\frac{L}{2},\ +\frac{L}{2}]$, is defined by:
\beq\label{a2} H \equiv -\frac{\hbar^2}{2m}D^2\ ,\quad  {\cal{D}}(H)= \
\left\{\phi,\,H\phi\ \in\ {\cal L}^2(-\frac{L}{2},
\,+\frac{L}{2})\ \ ,\quad\phi(\pm \frac{L}{2}) = 0\ \right\},
\eeq where $D$ is the differential operator $\dst\frac{d}{dx}\,$ and $\,{\cal
D}(H)\,$  is the definition domain of the operator $\,H.$

Two series of normalised eigenfunctions of opposite parity are obtained.  They
vanish outside the well and for 
$\,x \in [-\frac{L}{2}, +\frac{L}{2}]\,$ they write :
\beqa\label{a3} 
odd\ ones\ :\ & \dst\Phi_n(x) =\sqrt{\frac{2}{L}}\sin[
\frac{2n\pi x}{L}]\ ,& E_n = \frac{\hbar^2}{2m}\left(
\frac{2n\pi}{L}\right)^2 \nnb\\ even\ ones\ :\ & \dst\Psi_n(x)
=\sqrt{\frac{2}{L}}\cos[
\frac{(2n-1)\pi x}{L}]\ ,& E'_n = \frac{\hbar^2}{2m}\left(
\frac{(2n-1)\pi}{L}\right)^2 \ .
\eeqa where $n$ is a strictly positive integer.  The functions $\Phi_n(x)$ and
$\Psi_n(x)$ are continuous at 
$\,x =\pm \frac{L}{2}\,$ where they vanish. 

A question of fundamental  importance arises : is the Hamiltonian operator $H$ a
truly self-adjoint operator ?  To discuss more thoroughly this question let us
consider a particle  in the state defined by the even, normalised wave function :
\beq\label{a4}
\Psi(x) = -\sqrt{\frac{30}{L^5}}(x^2 - \frac{L^2}{4})\ ,\ 
\ |x|\le \frac{L}{2} \ ;\ \  \Psi(x) = 0\ ,\ 
\ |x|\ge \frac{L}{2}\ . 
\eeq
 It may be expanded \cite{expansion} on the complete basis of eigen functions of
$H$ given in  (\ref{a3}) :
\beq\label{a5}
\Psi(x) = \sum_{n=1}^{\infty}b_n \Psi_n(x)\ ,\ \ b_n  =(\Psi_n,\ \Psi) =
 \frac{(-1)^{n-1}}{(2n-1)^3}\frac{8\sqrt{15}}{\pi^3}\ .
\eeq

Let us define also, for further use,
\beq\label{a6}
\tilde{\Psi}(x) = -\frac{\hbar^2}{2m}D^2 \Psi(x) = 
\frac{\hbar^2}{m}\sqrt{\frac{30}{L^5}}\ ,\qq\quad -L/2<x<+L/2,
\eeq and let us  begin with some elementary computations : the mean value of the
energy and its mean-square deviation
 in the state (\ref{a4}). On the one hand we have
\beq\label{a8} <E> = \sum_{n=1}^{\infty} |b_n|^2E'_n = \frac{480\hbar^2}{m\pi^4
L^2}\sum_{n=1}^{\infty}
\frac{1}{(2n-1)^4} =  \frac{5\hbar^2}{mL^2},
\eeq but on the other hand 
$$<E> = (\Psi, H\Psi) = (\Psi, 
\tilde{\Psi}) = -\frac{30\hbar^2}{mL^5}\int_{-L/2}^{+L/2} [x^2-\frac{L^2}{4}]dx
=  \frac{5\hbar^2}{mL^2} =  \frac{10}{\pi^2}E'_1 \,.$$ These results are
coherent. Things are different for the energy mean-square  fluctuation. On the
one hand
\beq\label{a9} <E^2> = \sum_{n=1}^{\infty} |b_n|^2(E'_n)^2 = 
\frac{240\hbar^4}{m^2\pi^2 L^4}\sum_{n=1}^{\infty}\frac{1}{(2n-1)^2} =  
\frac{30\hbar^4}{m^2L^4},\eeq leads to
\beq\label{aa9}
\De E \equiv\sqrt{<E^2>-<E>^2}= \sqrt{5}\frac{\hbar^2}{mL^2},
\eeq and on the other hand 
$$<E^2> = (\Psi, H^2\Psi) = (\Psi, H\tilde{\Psi}) =  0\ \ !!$$ In order to 
understand the origin of the paradox, let us come back to the  definitions. The
probability of being in the eigenstate
$\phi_n$ of energy $\epsilon_n$ being given by $|(\phi_n,\Psi)|^2$, one  obtains
$$<E^2> = \sum_{n=1}^{\infty} \epsilon_n^2|(\phi_n,\Psi)|^2 =
\sum_{n=1}^{\infty} \epsilon_n^2(\phi_n,\Psi)(\Psi,\phi_n) = 
\sum_{n=1}^{\infty} (H\phi_n,\Psi)(\Psi,H\phi_n) $$ where the reality of the
eigenvalues of the Hamiltonian has been used. If H  were self-adjoint, one would
obtain with the help of the closedness relation
\beq\label{a10} <E^2> = \sum_{n=1}^{\infty} (\phi_n,H\Psi)(H\Psi,\phi_n) =
(H\Psi, H\Psi)
 =  (\tilde{\Psi},\tilde{\Psi}) = \frac{30\hbar^4}{m^2L^4}\,.
\eeq in agreement with the direct calculation (\ref{aa9}). But, if the
self-adjointness of $H$  was used once more, one would get
\beq\label{a11} <E^2> = (H\Psi, H\Psi) = (\Psi, H^2\Psi) = 0
\eeq which is necessarily wrong. In fact, in (\ref{a10}), we  used (correctly,
as shown by the  standard proof  using an integration by parts ) the
self-adjointness of $H$ when it acts in the set of functions that vanish at both
end-points  of the  well 
$$(H\phi_n,\Psi) = (\phi_n,H\Psi)\ \ \ ,\ \ \ (\Psi,H\phi_n) =
 (H\Psi,\phi_n)\ ;$$  
on the  contrary, in (\ref{a11}), the function
$\tilde{\Psi}$  does not belong  to that set and, consequently, in the 
integration by parts, the integrated term remains and
$$(H\Psi, \tilde{\Psi}) \neq (\Psi, H\tilde{\Psi})\,.$$

\noindent These simple calculations show that the problem lies in the 
definition of the action of the operator $H$ on a function $\tilde{\Psi}$  that
does not vanish at the end-points.

To summarise, we came up against the difficulty of the definition of a 
self-adjoint operator in a closed interval 
$[-L/2,\,+L/2]$ as an extension of a differential operator
$-\frac{\hbar^2}{2m}D^2$, question already solved by mathematicians in the 
thirties. Before explaining this theory in a  simple manner,  we analyse in the
next Section the momentum operator $-i\hbar D.$

\section{ Self-adjoint extensions of the momentum operator : a first approach}

Let us consider the one-dimensional momentum operator
$\,P = -i\hbar D\,$ in a closed $x$ interval.  Let us take for domain $\,{\cal
D}\,$ the following space
$${\cal{D}}(P) = \left\{\phi,\,\phi' \in {\cal{L}}^2([0,\,L])\ ;
\   \phi(0) = \phi(L) = 0\right\}.$$   The vanishing of  
\beqa\label{b1} (\psi,-i\hbar D\phi) - (-i\hbar D\psi,\phi) & =  & 
\int_{0}^{L}dx\left[ \ol{\psi}(x)(-i\hbar\frac{d\phi(x)}{dx}) -
(i\hbar\frac{d\ol{\psi}(x)}{dx})\phi(x)\right] =\nnb \\ = \
-i\hbar\int_{0}^{L}dx \frac{d}{dx} [\ol{\psi}(x)\phi(x)] & = & 
-i\hbar[\ol{\psi}(L)\phi(L) - 
\ol{\psi}(0)\phi(0)]\,.
\eeqa implies that $P$ is a symmetric operator in 
${\cal D}\,.$ But $P$ is not a self adjoint operator even if its adjoint 
$\,P^{\dg} = -i\hbar D\,$ has the same formal expression, 
\underline{but it acts on  a different space of functions.} Indeed, 
$${\cal{D}}(P^{\dg}) = \left\{\psi,\,\psi' \in {\cal{L}}^2([0,\,L])\ ;
\ {\rm no\ other\ restriction\ on}\ \psi(x)\right\}\,.$$  With (\ref{b1}), one
easily sees that the adjoint of the operator 
$\,P_{\la} = - i\hbar D\,$ acting on the subspace of ${\cal{L}}^2([0,\,L])$ 
such as 
$$\phi(L) = \la\phi(0)\ ,\ \ {\rm where}\ \la \in {\mb C}$$ is the operator
$P_{\la^{'}}$
 where $\la^{'} = 1/\ol{\la}\,.$ As a consequence, a candidate family of
self-adjoint extensions of the  operator $-i\hbar D$, depending on a complex
parameter
$\la \equiv 1/\ol{\la}\ ,\ i.\,e.\,$ a phase 
$\,\la=e^{i\tht},\ \ \tht\in[0,2\pi]$ is:
\beq\label{b4} P_{\tht} \phi(x) =
 -i\hbar D \phi(x)\ ,\quad{\cal D}(P_{\tht})=
\left\{ \phi,\,\phi'\in {\cal L}^2([0,\ L]),
\ \ \phi(L) = e^{i\tht}\phi(0)\ \ \right\}\eeq Notice that for $\,\tht=0\,,$ one
recovers the usual periodic boundary conditions.

\underline{Conclusion} : A symmetric differential operator acting on a 
given functional space is not automatically a self-adjoint operator and  may
have none, a unique or an infinity of self-adjoint extensions. In the next  Section,
we give some mathematical results on the theory of self-adjoint  extensions of a
differential operator in a Hilbert space and  ``deficiency indices".

\section{Deficiency indices and von Neumann's theorem} Since this Section makes
use of mathematical terminology, let us  begin with some precise definitions.

Let us consider a Hilbert space $\,{\cal H}.$ An operator $\,(A,{\cal D}(A))\,$
defined on $\,{\cal H}\,$ is said to be densely defined if the subset 
$\,{\cal D}(A)\,$ is dense in $\,{\cal H},$ i.e. that for any 
$\,\psi\in\,{\cal H}\,$ one can find in $\,{\cal D}(A)\,$ a sequence 
$\,\phi_n\,$ which converges in norm to $\,\psi.$

An operator $\,(A,{\cal D}(A))\,$ is said to be {\em closed} if $\,\phi_n\,$  is
a sequence in $\,{\cal D}(A)\,$ such that
$$\dst\lim_{n\to\nf}\phi_n=\phi,\qq\qq\lim_{n\to\nf}A\phi_n=\psi,$$ then
$\,\phi\in{\cal D}(A)\,$ and $\,A\phi=\psi.$

Let us recall the definition of the adjoint operator of a (in general not 
bounded) operator $\,H\,$ with dense domain $\,{\cal D}(H).$ The domain 
$\,{\cal D}(H^{\dg})\,$ is the space of functions $\,\psi\,$ such that  the
linear form
$$\phi\quad\longrightarrow\quad (\psi,H\phi)$$ is continuous for the norm of
$\,{\cal H}.$ Hence there exists a 
$\psi^{\dg}\in{\cal H}$ such that
$$(\psi,H\phi)=(\psi^{\dg},\phi).$$ One defines $\,H^{\dg}\psi=\psi^{\dg}.$ A
useful result is that the adjoint of any densely defined operator is closed, 
see \cite[p. 80, vol. 1]{AG}.

An operator $\,(H,{\cal D}(H))\,$ is said to be {\em symmetric} if for all 
$\,\phi,\psi\in{\cal D}(H)\,$ we have
$$(H\phi,\psi)=(\phi,H\psi).$$ If $\,{\cal D}(H)\,$ is dense, it amounts to
saying that 
$\,(H^{\dg},{\cal D}(H))\,$ is an extension of $\,(H,{\cal D}(H)).$

The operator $\,H\,$ with dense domain $\,{\cal D}(H)\,$ is said to be 
self-adjoint if $\,{\cal D}(H^{\dg})={\cal D}(H)\,$ and $\,H^{\dg}=H.$

In this Section we will assume that $\,(A,{\cal D}(A))\,$ is densely defined, 
symmetric and closed and let $(A^{\dg},{\cal D}(A^{\dg}))$ be its adjoint.

One defines the deficiency subspaces ${\cal N}_{\pm}$ by
$$\barr{l} {\cal N}_+=\{\psi\in{\cal D}(A^{\dg})\ , \qq A^{\dg}\psi=z_+\psi,
\qq \mbox{Im}\,z_+>0\}, \\ [3mm] {\cal N}_-=\{\psi\in{\cal D}(A^{\dg})\ , \qq
A^{\dg}\psi=z_-\psi,
\qq \mbox{Im}\,z_-<0\},\earr$$ with respective dimensions $n_+,n_-.$ These are
called the deficiency  indices of the operator $A$ and will be denoted by the
ordered  pair $(n_+,n_-).$

The crucial point is that $n_+$ (resp. $n_-$) is completely independent of  the
choice of $z_+$ (resp. $z_-$) as far as it lies in the upper  (resp. lower)
half-plane. It follows that a simple way to determine 
$(n_+,n_-)$ is to take $z_+=i\la$ and $z_-=-i\la$ with an arbitrary strictly 
positive constant $\,\la\,$ needed for dimensional reasons. 

The following theorem, first discovered by Weyl \cite{We} in 1910 for second 
order differential operators and generalized by von Neumann \cite{vN} in  1929,
is of primary importance
\begin{theorem} For an operator $\,A\,$ with deficiency indices $\,(n_+,n_-)\,$
there are  three possibilities :
\brm
\item If $n_+=n_-=0,$ then $A$ is self-adjoint (in fact this is a necessary  and
sufficient condition).
\item If $n_+=n_-=n\geq 1,$ then $A$ has infinitely many  self-adjoint
extensions, parametrized by a unitary $n\times n$ matrix  (i. e. $n^2$ real
parameters).
\item If $n_+\neq n_-,$ then $A$ has no self-adjoint extension.
\erm\end{theorem}

The application of this theorem to differential operators requires still a  lot
of work : even if we start from an operator $P$ which is formally  self-adjoint,
this does not prove that $P$ is truly  self-adjoint because the domains ${\cal
D}(P)$ and ${\cal D}(P^{\dg})$ will  be different in general.

For a given differential operator $P$ one has to solve three problems :
\brm
\item Find a domain ${\cal D}(P)$ for which the formally self-adjoint  operator
$P$ is symmetric and closed.
\item Compute its adjoint $(P^{\dg},{\cal D}(P^{\dg}))$ and determine  the deficiency
indices of $P^{\dg}$.
\item When they do exist, describe the domains of all the self-adjoint 
extensions.
\erm

A whole body of theory has been built up to solve these problems and is  given
in many text-books (for instance \cite{AG},\cite{Na}).  In the next Section we
describe the results for the simplest case of the  momentum operator $P=-i\hbar
D$, referring for the proofs to 
\cite[vol. 1, p. 106-111]{AG}.

\section{Self-adjoint extensions of the momentum operator} Let us apply
the previous analysis to the momentum operator $P=-i\hbar D\,,$ in three
different  ``physical" situations : first on the whole real axis and in this
case we  conclude to a unique self-adjoint extension, second on the positive
semi-axis  and in this case there is no self-adjoint extension, and third in a
finite  interval $[0,L]$ in which case there are infinitely many self-adjoint 
extensions, parametrized by $U(1)\,,$ i.e. a phase. The momentum operator is
certainly the simplest differential operator to  begin with and it already
exhibits all the possibilities described in  von Neumann's theorem. For each
physical situation corresponding to position space being some interval (a,b), finite or
not, the maximal domain  on which the operator 
$P=-i\hbar D$ has a well defined action will be called 
${\cal D}_{\rm max}(a,b)$. In this Section, we apply the previous theorem, postponing some mathematical details to the
Appendix A.  

Let us consider the Hilbert space ${\cal H}=L^2(a,b)$ and to use von Neumann's
theorem, we have to determine the  functions $\psi_{\pm}(x)$ given by 
$$P^{\dg}\psi_{\pm}(x)=-i\hbar D\psi_{\pm}(x)=
\pm i\frac{\hbar}{d}\psi_{\pm}(x).$$ For dimensional reasons we have introduced
the constant $d>0$,  homogeneous to some length.

An easy integration gives $\psi_{\pm}(x)=C_{\pm}e^{\mp x/d}\,.$ Then we have to
discuss the different intervals (a, b).
 
\subsection{The operator $\,P\,$ on the whole real axis}
None of  the
functions $\psi_{\pm}(x)$ belong to the Hilbert space $L^2({\mb R})$ and
therefore the deficiency  indices are
$(0,0)\,.$ 
Hence we conclude that the operator $\,(P,{\cal D}_{\rm max}({\mb R}))\,$ is 
indeed self-adjoint, in agreement with the heuristic considerations given in 
the standard textbooks on quantum mechanics. Moreover, the spectrum of $P$ on the
real axis is  continuous, with no eigenvalues.

\subsection{The operator $\,P\,$ on the positive semi-axis} Among the functions
$\psi_{\pm}(x)\,,$  only $\psi_+$ belongs to
$L^2(0,+\nf)\,.$ We conclude to the  deficiency indices
$(1,0)$ and therefore, by the von Neumann theorem, 
$P$ has no self-adjoint extension. This is a fairly 
surprising conclusion, since it implies that the momentum is not a 
measurable quantity in that situation !

\subsection{The operator $\,P\,$ on a finite interval}

Since we are working on a finite interval, both $\psi_{\pm}(x)= C_{\pm}e^{\mp
x/d}$ belong to $L^2(0,L)$ and the deficiency indices are 
$(1,1)$. 

From von Neumann's theorem, we know that the self-adjoint extensions are 
parametrized by $U(1)\,,$ i.e. a phase $e^{i\tht}$, in agreement with the result 
of section 3. Denoting these extensions  by $P_{\tht}=(P,{\cal D}_{\tht})$, they
are given by
\beq\label{sa1}{\cal D}_{\tht}=\{\psi\in{\cal D}_{\rm max}(0,L),\quad \psi(L)=
e^{i\tht}\psi(0)\},\qq\tht\in[0,2\pi].\eeq

Moreover, the spectra  are  purely discrete. Using
the boundary condition (\ref{sa1}), the eigenvalues  and eigenfunctions are
easily shown to be
\beq\label{e1}
\left\{\barr{l}
\dst P_{\tht}\phi_n(x,\tht)=\frac{2\pi\hbar}{L}\nu 
\phi_n(x,\tht),\qq \nu=n+\frac{\tht}{2\pi},\qq n=0,\pm 1,\pm 2,\ldots \\ [3mm]
\dst \phi_n(x,\tht)=\frac 1{\sqrt{L}}\exp\left[2i\pi\nu\frac xL
\right],\qq (\phi_m,\phi_n)=\de_{mn}.\earr\right.
\eeq As the phase $\tht$ appears in the eigenfunctions any measurement of the 
momentum of a given system should, in general, depend on it. To display this, 
let us go back to the state (\ref{a4}). After a translation, we  are left with
the wave function
$$\Psi(x)=\sqrt{\frac{30}{L^5}}x(L-x).$$ Its eigenfunction expansion is 
$$\dst\Psi(x)=\sum_{n=-\nf}^{n=+\nf}c_n(\tht)\phi_n(x,\tht),$$ with
coefficients 
\beq\label{coef1} c_n(\tht)=-\frac{\sqrt{30}}{2\pi^2\nu^2}\left[\cos(\tht/2)-
\frac{\sin(\tht/2)}{\pi\nu}\right]e^{-i\tht/2},\qq{\rm for}\qq\tht\neq\, 0,\eeq
and
\beq\label{coef2} c_0=\frac{\sqrt{30}}{6},\qq\qq
c_n=-\frac{\sqrt{30}}{2\pi^2n^2},\qq n=1,2,\ldots
\qq{\rm for}\qq\tht= 0.\eeq So the probability to find the particle with a
momentum 
$\dst \frac{2\pi\nu\hbar}{L}$, being equal to $|c_n(\tht)|^2\,,$ is really 
$\tht$ dependent. Of course one would like to have a physical argument  which
gives some preferred value of $\tht$.

Let us conclude with the following remarks :
\brm
\item The textbooks which do study the momentum operator in a box (\cite{Ba} and
\cite[vol. 2, p. 1202]{CDL}), usually consider (using physical arguments) only
the  self-adjoint extension corresponding to the periodic boundary condition 
(i.e. $\tht=0$) which is certainly the simplest (but still arbitrary) choice. The anti-periodic boundary condition (i.e. $\tht=\pi$) has been 
considered by Capri in \cite{Ca}.

\item For a particle in a box, it is often argued that the ``physical" wave 
function should continuously vanish on the walls $\,x=0\,$ and $\,x=L,$ 
ensuring that the presence probability vanishes continuously for $\,x\leq 0\,$ 
and for $\,x\geq L.$ One should realize that the continuity of the measurable 
quantity
$${\rm Pr}(0\leq x\leq u)=\int_0^u\,|\phi(x)|^2\,dx,\qq\qq u\in[0,L]$$ is
ensured as soon as the integral $\,\dst\int_0^L\,|\phi(x)|^2\,dx\,$ does 
converge and does not require any  continuity property of $\,\phi(x).$
Specializing this remark to the  eigenfunctions of $\,P_{\tht}\,$ we observe
that $\,|\phi_n(x,\tht)|^2\,$ does not  vanish continuously at $\,x=0\,$ but
nevertheless the physical quantity
$${\rm Pr}(0\leq x\leq u)=\frac uL$$ vanishes continuously, as it should, for
$u\to 0.$

\item The existence of normalisable eigenfunctions of the momentum operator  has
an important consequence : the Heisenberg inequality 
$\,\De X\cdot\De P\geq\hbar/2$ no longer holds. Indeed, for the state 
$\,\phi_n(x,\tht)\,$ given in relation (\ref{e1}), one has $\,\De P=0\,$ and 
$\,\De X=L/2.$ On the contrary, on the whole real axis the spectrum is fully 
continuous (no normalisable eigenfunctions), and the momentum probabilities  are
related to the Fourier transformed wave function. As the widths in x-space  and
in p-space are inversely proportional, the Heisenberg inequality follows.

\item If one identifies the variable $\,x\,$ with the angular variable 
$\,\vp\in\,[0,2\pi]\,$ of polar coordinates, then the angular momentum is 
$\,L_z=-i\hbar\dst\frac{d}{d\vp}.$ The previous remark shows that the inequality 
$\,\Delta\vp\cdot\Delta L_z\geq\hbar/2\,$ can be violated, even by wave
functions  periodic in the angle $\,\vp.$

\erm

\section{Self-adjoint extensions of the Hamiltonian}  In the same  setting as in
the previous section, we consider now the
Hamiltonian operator $\,H=-D^2\,$. We work in the Hilbert space $\,L^2(a,b).$ The
maximal domain in which  the operator $\,D^2\,$ is defined will again be called 
$\,{\cal D}_{\rm max}(a,b).$
To compute the deficiency indices we solve
\beq\label{h1}-D^2\phi(x)=\pm i k_0^2\,\phi(x),\qq k_0>0\,,\eeq  and get 
\beq\label{h2}
\dst\phi_{\pm}=a_{\pm}e^{k_{\pm}x}+b_{\pm}e^{-k_{\pm}x},\qq k_{\pm}=\frac{(1\mp
i)}{\sqrt{2}}k_0.\eeq

\subsection{The Hamiltonian on the whole real axis} The physical situation
corresponds to a free particle moving in a one dimensional space.  The Hilbert space is
$\,{\cal H}=L^2({\mb R})\,$ which implies  
$\,\phi_{\pm}\not\in{\cal H}$ and the deficiency indices $(0,0).$ It follows 
that on the real axis there is a {\em unique} self-adjoint extension of  the
Hamiltonian, with a fully continuous spectrum, in full agreement with the 
physicist understanding of this case.

\subsection{The Hamiltonian on the positive semi-axis} The physical problem is
that of a free particle in front of an infinitely  high wall for $\,x<0.$ In the
Hilbert space ${\cal H}=L^2(0,+\nf)$ we have  the solutions to equation
(\ref{h1}) given by
$$\phi_{\pm}=b_{\pm}e^{-k_0 x/\sqrt{2}}e^{\pm ik_0 x/\sqrt{2}},$$ leading to the
deficiency indices $(1,1),$ and therefore to infinitely many  self-adjoint
extensions parametrized by $\,U(1).$

The corresponding boundary conditions are
$$(\phi'(0)-i\phi(0))=e^{i\alf}(\phi'(0)+i\phi(0)),
\qq\quad\alf\in\,[0,2\pi],$$
which are equivalent to
\beq\label{h3}
\phi(0)=\la\phi'(0),\qq\qq\la = -\tan(\alf/2),\qq\qq\la\in{\mb R}\cup\{\nf\},\eeq see 
\cite[vol. 2, p.187, 204]{AG}. The boundary condition 
$\,\phi'(0)=0\,$ corresponds to $\la=\nf\,.$ Physicists use the 
particular extension with $\,\la=0,$ see for instance \cite[p. 328]{JMLL}  and \cite[p. 33]{Sch}.

Let us now discuss the energy-spectra of a particle confined in the region
$\,x\geq 0\,.$ When the particle energy $\,E\,$ is positive, we can compute the
reflexion  coefficient for this infinitely high barrier in order to compare the 
predictions given by the different extensions. The wave function is
\beq\label{hp1}
\dst \phi(x)=A\, e^{-ikx}+B\,e^{ikx},\qq\qq E=\frac{\hbar^2k^2}{2m},\qq k>0.\eeq
Let us define the reflection amplitude and reflection probability by
$$r(k)=\frac AB,\qq\qq\qq R(k)=|r(k)|^2.$$ Imposing the boundary condition
(\ref{h3}) we get
\beq\label{hp2} r(k)=-\frac{1+i\la k}{1-i\la k}\qq\Rightarrow\qq R=1.\eeq
Remarkably enough the physical content (i.e. $\,R=1$ !) of all the extensions 
is the same : the wall acts as a perfect reflector.

This is not quite true for the bound states
$$E=-\frac{\hbar^2\rho^2}{2m},\qq \rho>0,\qq\qq\phi(x)=A\,e^{-\rho x},$$ for
which (\ref{h3}) implies $\,(1+\la\rho)A=0.$ There will be a bound state  with
$\,\rho=-1/\la\,$ only for $\,\la<0\,$ and different from $\,\nf.$  Its energy
and normalised wave function are
\beq\label{hp3}
\dst E=-\frac{\hbar^2}{2m\la^2},\qq\la<0,\qq \qq 
\phi(x)=\sqrt{\frac 2{|\la|}}e^{-x/|\la|}.\eeq As far as an infinitely high wall
is feasible experimentally, the  existence (or non-existence) of this negative
energy will act as a selector  of some self-adjoint extensions.

If experiment rules out the negative energy state, or  if one is reluctant to
accept negative energies for the Hamiltonian,  there are still many possible
extensions, with $\,
\la\geq 0\,$ or $\,\la=\nf.$

In an attempt to lift this degeneracy, we consider the simplified  deuteron theory described by the potential
\beq\label{hp4} V(x)=\left\{\barr{l}
\nf\qq{\rm for}\quad x<0,\\[3mm] -V_0\qq{\rm for}\quad 0<x<a,\qq V_0>0,\\[3mm]
\ 0\qq{\rm for}\quad x>a.\earr\right.\eeq The wave function is well known to be
\beq\label{hp5}\barr{l}
\dst x<a\ :\qq\phi_1(x)=A\,\sin kx+B\,\cos kx,\qq\qq 
E+V_0=\frac{\hbar^2k^2}{2m},\ \,\quad k>0,\\[3mm]
\dst x>a\ :\qq\phi_2(x)=C\,e^{-\rho x},\qq\qq\qq\qq\qq\qq 
E=-\frac{\hbar^2\rho^2}{2m},\quad \rho>0.\earr
\eeq We next impose the boundary condition (\ref{h3}) and the usual continuity 
conditions at $\,x=a.$ Using the notations $\,X=ka\,$ and $\,Y=\rho a\,,$ we 
get that the bound state energy is given by the solution of the system
\beq\label{hp6}
\dst \la\geq 0  \qq\quad\longrightarrow\quad   Y=-X\frac{1-(\la/a)
X\tan\,X}{\tan\, X+(\la/a) X},
\qq{\rm and}\qq \frac{V_0}{|E|}=1+\left(\frac XY\right)^2.\eeq In the case of
the deuteron, the absolute value of the binding energy $\,|E|,$  is roughly
equal to $\,2.2\,$ MeV. Its size is $\,a=2\,$ F and we take  
$\,2m=M\,$ where $\,M\,$ is the nucleon mass. It follows that $\,Y=0.46.$ For a 
given value of $\,\la,$ we have to solve for $\,X,$ and then  recover the
potential $\,V_0.$ Numerical analysis gives the following dependence  on
$\,V_0\,$ with respect to the parameter $\,\la$ :

\vspace{5mm}
\centerline{
\begin{tabular}{|c||cc|c|c|c|c|c|c|c|c|c|c|}
\hline
$\la/a$ & & 0 & 0.1 & 0.2 & 0.5 & 1 & 2 & 5  & 10 & 100 & $\nf$ \\
\hline
$V_0\ $(MeV) &  & 36.8 & 31.5 & 27.5 & 20.5 & 15.3 & 11.5 & 8.59   & 7.50  &
6.47 & 6.34  \\
\hline
\end{tabular}}

\vspace{5mm} Let us observe that the parameter $\,\la,$ describing the different
extensions,  does indeed have an effect on physical quantities (as already
observed for  the momentum operator, in subsection 5.3) and in fact experiment,
not just  theoretical prejudices, should decide which is the ``right" value for
it.

\subsection{The Hamiltonian on a finite interval} This last case corresponds to
a particle in a box : $x\in[0,L].$ From a  mathematical standpoint the situation
is quite similar to the one already  experienced with the momentum operator in
the previous section, but up to our knowledge, it did not appear before in the
literature. So we give some details in the main text.

One starts from the operator $\,(H,{\cal D}_0(H))\,$ such that
$${\cal D}_0(H)=\{\phi\in{\cal D}_{\rm max}(0,L)\quad{\rm
 and}\quad\phi(0)=\phi(L)=\phi'(0)=\phi'(L)=0.\}$$ It is densely defined and
closed, with adjoint
$$H^{\dg}=H,\qq{\cal D}(H^{\dg})={\cal D}_{\rm max}(0,L).$$ Since all the
solutions of equation (\ref{h1}) belong to $\,L^2(0,L),$ the deficiency indices
are now $\,(2,2)\,$ and the self-adjoint extensions are  parametrized by a
$\,U(2)\,$ matrix.

To describe these self-adjoint extensions, it is natural to introduce the 
sesquilinear form, for $\,\phi\,$ and $\,\psi\,$ in $\,{\cal D}_{\rm max}(0,L),$
\beq\label{oublieh3}
\dst B(\phi,\psi)=\frac 1{2i}\left((H^{\dg}\phi,\psi)-(\phi,H^{\dg}\psi)\right)
\eeq which depends only on the boundary values of $\,\phi\,$ and $\,\psi.$
Specializing  to $\,\psi=\phi\,$ we have
\beq\label{h4}\dst  B(\phi,\phi)=\frac
1{2i}\left(\phi'(L)\ol{\phi(L)}-\phi(L)\ol{\phi'(L)}
-\phi'(0)\ol{\phi(0)}+\phi(0)\ol{\phi'(0)}\right).\eeq The identity
\beq\label{id}
\dst \frac 1{2i}\left(x\ol{y}-\ol{x} y\right)=
\frac 14\left(|x+iy|^2-|x-iy|^2\right),\eeq applied to $\, x=L\phi'(L),\
y=\phi(L)\,$ and $\,x=L\phi'(0),\ y=\phi(0)\,$  brings relation (\ref{h4}) to
\beq\label{h5} 4LB(\phi,\phi)=|L\phi'(0)-i\phi(0)|^2+|L\phi'(L)+i\phi(L)|^2
-|L\phi'(0)+i\phi(0)|^2-|L\phi'(L)-i\phi(L)|^2.
\eeq The domain of a self-adjoint extension is a maximal subspace of 
$\,{\cal D}_{\rm max}(0,L)\,$ on which the form $\,B(\phi,\phi)\,$ vanishes 
identically. These self-adjoint extensions are parametrized by a unitary  matrix
$\,U,$ and will be denoted $\,H_U=(H,{\cal D}(U)),$ in which 
$\,{\cal D}(U)\,$ is the space of functions $\,\phi\,$ in 
$\,{\cal D}_{\rm max}(0,L)\,$ satisfying the following boundary conditions 

\beq\label{r1}
\dst \left(\barr{l} L\phi'(0)-i\phi(0) \\ [2mm]  L\phi'(L)+i\phi(L)
\earr\right)=U
\left(\barr{l} L\phi'(0)+i\phi(0) \\ [2mm]  L\phi'(L)-i\phi(L) \earr\right).\eeq
Notice the arbitrariness in the choice of the ordering of the  coordinates
$\,L\phi'(0)\pm i\phi(0)\,$ and $\,L\phi'(L)\mp i\phi(L).$ The  crucial
observation is that whatever the choice of coordinates is, the  arbitrariness of
the self-adjoint extensions remains described by a 
$\,U(2)\,$ matrix.

These boundary conditions describe all the self-adjoint extensions 
$H_U=(H,{\cal D}(U))$ of a particle in a box.  Moreover, thanks to the useful
theorem,  proved in
\cite[vol. 2, p. 90]{Na}, stating that for a differential  operator of order
$\,n\,$ with deficiency indices $\,(n,n)\,$ all of its  self-adjoint extensions
have a discrete spectrum, we know that all the  spectra of the $\,H_U\,$ are
fully discrete. Leaving the details of these spectra to the Appendix B, we only give the results.

Let us parametrize the unitary matrix $U$ as : 
\beq\label{r10}
\dst U=e^{i\psi}\ M,\qq\det M=1,\qq\Rightarrow\qq 
\det U=e^{2i\psi},\qq\psi\in\,[0,\pi]\qq \eeq where $M$ is an element of
$\,SU(2),$ i.e. a unitary matrix of  determinant $\,1.$ The range of $\psi$ is
restricted to $\pi$ instead  of $2\pi$ because the couples $(\psi,\ M)$ and
$(\psi+\pi,\ -M)$ give  rise to the same unitary matrix $U$. Notice also that it
follows that  the points $\,\psi=0\,$ and $\,\psi=\pi\,$ are to be identified.

To parametrize the matrix  $M\,,$ we used the  Pauli matrices 
$$\tau_1=\left(\barr{cc} 0 & 1\\[3mm] 1 & 0\earr\right),\quad
\tau_2=\left(\barr{cc} 0 & -i\\[3mm] i & 0\earr\right),\quad
\tau_3=\left(\barr{cc} 1 & 0\\[3mm] 0 & -1\earr\right),$$ and the  notation :
$\quad\dst\vec{n}\cdot\vec{\tau}=n_1\,\tau_1+n_2\,\tau_2+n_3\,\tau_3\,.\quad$
With coordinates $m=(m_0,\vec{m})$ constrained by
\beq\label{r13} m_0^2+\vec{m}\cdot\vec{m}=1\qq\Longleftrightarrow\qq m\in{\mb
S}^3,\eeq
$M$ writes :  
\beq\label{r12} M=\left(\barr{cc} m_0-im_3 & -m_2-im_1 \\ [3mm] m_2-im_1 &
m_0+im_3\earr\right)=m_0\,I-i\,\vec{m}\cdot\vec{\tau}.\eeq

\noindent Then, starting from the boundary
conditions (\ref{r1}), we obtain the spectra for the Hamiltonian in a box (see
details in Appendix B) :
\beq\label{r14}
\barr{l}
\hspace{-2cm} a)\ \dst E=\frac{s^2}{L^2} >0\,  : \\[3mm]
\hspace{2cm}2s[\sin\psi\cos s-m_1] =  \sin s[\cos\psi(s^2+1)-m_0(s^2-1)], \\[3mm]
\hspace{-2cm} b)\ \dst E=0\,: \hspace{1.8cm} s \rightarrow 0\ {\rm in\ result\ (\ref{r14}-a)}  \Leftrightarrow \\[3mm]
 \hspace{9mm}\hspace{2cm} 2\sin\psi-\cos\psi =  2m_1+m_0,  \\
\hspace{-2cm} c)\ \dst E=-\frac{r^2}{L^2} <0\,:  \quad s = 
ir\ {\rm in\ result\ (\ref{r14}-a)} \Leftrightarrow    \\[3mm]
\hspace{2cm} 2r[\sin\psi\,\cosh r-m_1]\hspace{1mm}  =  
\sinh r[-\cos\,\psi(r^2-1)+m_0(r^2+1)].\earr
\eeq

\noindent{\bf Remarks :}
\brm
\item The eigenvalue equations are independent of the parameters 
$(m_2,\,m_3).$ As shown in appendix B, this follows from their invariance  under the transformation
$$ M\quad\to\quad  M'=e^{-\tht\tau_1/2i}\,M\,e^{+\tht\tau_1/2i}.$$
Let us  point out that this invariance is specific of 
the spectra, not of the eigenfunctions.

\item The existence of negative  energies seems rather surprising since $P^2=-D^2$ is
a formally  positive operator. That this is not generally true can be seen by
computing
$$(\phi,H\phi)-(P\phi,P\phi)=\ol{\phi}(0)\phi'(0)-\ol{\phi}(L)\phi'(L),
\qq\phi\in D_U.$$ If the right hand side of this relation is positive, then the
spectrum  will be positive, an issue which depends on the extension $H_U$
considered (see section 7.3).
\erm 

\section{Restrictions from physics on the self-adjoint extensions} In the
previous section we have described all the possible self-adjoint  extensions of
the operator $\,H_U\,$ as they follow from operator theory.  Now we examine
which extensions are likely to play an interesting role  according to arguments
from physics.

\subsection{Extensions preserving time reversal} Let $\Psi(x,t)$ be a solution
of the Schr\"odinger equation
\beq\label{t1}
\dst i\hbar\frac{\pt\Psi}{\pt t}(x,t)= -\frac{\hbar^2}{2m}\frac{\pt^2\Psi}{\pt
x^2}(x,t)\eeq inside the box. The time reversal invariance of this equation
means that if 
$\Psi(x,t)$ is a solution of (\ref{t1}), then $\ol{\Psi}(x,t)$ is also a 
solution. If we consider a stationary state of definite energy $\,E\,$ with  the
wave function
$$\dst\Psi(x,t)=\phi_E(x)\,e^{-i\frac{Et}{\hbar}},$$ the previous statement
implies that $\phi_E(x)$ and $\ol{\phi}_E(x)$ are two  eigenfunctions of the
Hamiltonian $H$ with the same eigenvalue $E.$ One can  therefore choose {\em
real} eigenfunctions by taking the linear combination 
$\,\phi_E(x)+\ol{\phi}_E(x).$ 

The shortcoming in this argument is that the boundary conditions (\ref{r1})  do
not lead {\em necessarily} to real eigenfunctions $\phi_E(x).$  Among all of the
self-adjoint extensions of the Hamiltonian only some subclass  will have real
eigenfunctions. These extensions will be said to be time  reversal invariant.

To determine all of these extensions, we merely observe that, using the notations
$$\psi_{\pm}(x)=L\phi'(x)\pm i\phi(x),$$  the reality of $\,\phi(x)\,$ implies
$\,\ol{\psi}_{\pm}(x)=\psi_{\mp}(x).$  Taking the complex conjugate of relation
(\ref{r1}) gives
\beq\label{t2}\dst \left(\barr{l}
\psi_+(0) \\ [2mm]  \psi_-(L) \earr\right)=\ol{U}
\left(\barr{l}
\psi_-(0) \\ [2mm]  \psi_+(L) \earr\right)=\ol{U}\, U
\left(\barr{l}
\psi_+(0) \\ [2mm]  \psi_-(L) \earr\right).\eeq Since $\,\psi_+(0)\,$ and
$\,\psi_-(L)\,$ cannot vanish simultaneously,  we conclude to
\beq\label{t3}
\det({\mb I}-\ol{U}\, U)=0.\eeq Using for $\,U\,$ the coordinates given by
(\ref{r12}), easy computations give $\,m_2=0$ and, correspondingly, the
matrix
\beq\label{t4} U=e^{i\psi}\left(\barr{cc} m_0-im_3 & -im_1 \\[3mm] -im_1 &
m_0+im_3\earr\right)\quad{\rm with}\quad \psi\in[0,\pi]
\quad{\rm and}\quad m_0^2+m_1^2+m_3^2=1,\eeq

\subsection{Extensions preserving parity} The potential $V(x),$ vanishing inside
the box, is symmetric with respect  to the point $x=L/2.$ To make this symmetry
explicit we shift the  coordinate $x$ to
$$u=\frac xL-\frac 12,\qq\qq u\in[-\frac 12,+\frac 12],$$ and define
$$\tilde{V}(u)=V(x),\qq\qq \tilde{\phi}_E(u)=\phi_E(x).$$ In the new variable
$\,u\,$ the potential is even : 
$\,\tilde{V}(-u)=\tilde{V}(u).$  It follows that, for a given energy, the
eigenfunctions 
$\,\tilde{\phi}_E(u)\,$ and $\,\tilde{\phi}_E(-u)\,$ are solutions of the  same
differential equation and we can choose linear combinations of definite  parity
$\,\tilde{\phi}_E(u)\pm \tilde{\phi}_E(-u).$

As was already the case in the discussion of time reversal invariance, this 
argument is wrong since it overlooks the possibility for the boundary 
conditions (\ref{r1}) to break parity. Note that this point is often forgotten
in Quantum Mechanics textbooks : there, one generally finds that, as soon as the
potential is symmetric, the solution of the Schrodinger equation is 
of definite parity. It should be clear that the boundary conditions 
are essential. 
A good example to think about is the finite square well. The wave 
functions of its bound states are subject to the 
boundary condition $\dst\int\,|\phi(x)|^2\,dx<\nf.$ As this condition 
is symmetric, the wave functions do have a definite 
parity. This is not the case for the diffusion eigenfunctions, for which 
one has an incoming and reflected wave for $x\to -\nf,$ while for 
$x\to +\nf$ one has only a transmitted wave. In this second case the 
symmetry between $x$ and $-x$ is broken by the very conditions 
which characterize a diffusion experiment.

We will therefore define parity preserving extensions of the Hamiltonian 
$\,H_U\,$ as the ones for which the eigenfunctions 
$\,\tilde{\phi}_E(u)\,$ verify
\beq\label{p1} |\tilde{\phi}_E(-u)|^2=|\tilde{\phi}_E(u)|^2.\eeq Here one finds
(Appendix C) that all parity preserving extensions are given by 
$\,m_3=0\,$ and so correspond to the matrix
\beq\label{p5} U=e^{i\psi}\left(\barr{cc} m_0 & -m_2-im_1\\[3mm] m_2-im_1 &
m_0\earr\right),\qq\psi\in[0,\pi],\quad m_0^2+m_1^2+m_2^2=1.\eeq

\subsection{Extensions preserving positivity} One of the most surprising facts,
for a physicist, is the appearance of  extensions with negative energies (these
can be determined explicitly in  some particular cases, see appendix B).

 From a theorem proved in \cite[theorem 16, vol. 2, p. 44]{Na} one knows that 
only a finite number of negative energies can appear and that the sum of their 
multiplicities is at most 2. However, the determination of the $U$ matrices 
with no negative eigenvalues, involves lengthy graphical discussions of 
equation (\ref{r14}), which are fairly tedious.

A partial answer to this problem is offered by an interesting theorem due to 
von Neumann (see \cite[p. 97]{AG}). It states that if $A$ is densely  defined
and closed, then $\,A^{\dg}A\,$ is self-adjoint (and obviously positive).

Let us apply this result to the operator $(P=-iD,{\cal D}_0(P))$ defined in 
Subsection 5.3, whose adjoint was $(P,{\cal D}_{\rm max}(0,L)).$ It follows 
that the operator
$$(P^2,{\cal D}_1(P^2)),\qq{\cal D}_1(P^2)=
\{\phi\in{\cal D}_{\rm max}(0,L)\ \ {\rm with}\ \ \phi(0)=\phi(L)=0\},$$ will be
self-adjoint. It does correspond to the extension with $\,U={\mb I}.$

If we take for operator $(P,{\cal D}_{\rm max}(0,L)),$ with adjoint 
$(P,{\cal D}_0(P))$, we are led to
$$(P^2,{\cal D}_2(P^2)),\qq{\cal D}_2(P^2)=
\{\phi\in{\cal D}_{\rm max}(0,L)\ \ {\rm with}\ \ \phi'(0)=\phi'(L)=0\},$$ a
self-adjoint extension corresponding to $\,U=-{\mb I}.$

As a last example, we may start from $(P,{\cal D}_{\tht}),$ in which case  von
Neumann's theorem gives the self-adjoint extension
$$(P^2,{\cal D}_3(P^2)),\qq{\cal D}_3(P^2)=
\{\phi\in{\cal D}_{\rm max}(0,L)\ \ {\rm with}\ \ \phi(L)=e^{i\tht}\phi(0),
\quad \phi'(L)=e^{i\tht}\phi'(0)\},$$ corresponding to the matrix
$$U=\left(\barr{cc} 0 & e^{-i\tht}\\[3mm] e^{i\tht} &
0\earr\right),\qq\qq\tht\in[0,2\pi].$$ As shown in the appendix B.2., for
this choice of matrix $\,U,$ the operators 
$\,(P^2,{\cal D}_U)\,$ and $\,(P,{\cal D}_{\tht})\,$ have the same
eigenfunctions. These extensions, $\,(P^2,{\cal D}_U)\,$ are really the square
of the ones of the momentum operator $\,(P,{\cal D}_{\tht})\,.$

\subsection{The infinite well as a limit of the finite one.}

Let us consider the standard problem of a particle of mass  $m$ in a one 
dimensional potential well of  width $L$ and depth $V_0\ $:
\beq\label{f0} V(x) = 0\ ,\ \ x\in\ ]0,\ L[\ \ ;\ \  \ \ \ V(x) =  V_0\ \ >\ 0\
,\ \ x\,\not\in\,]0,L[.
\eeq A standard computation gives the bound states wave function
\beq\label{f1}\barr{cllc} x\ \le 0\ \ :  & \dst\phi_n(x) = & d_n\,e^{\rho x} &
\dst\quad \rho^2  = 
\frac{2m(V_0 - E)}{\hbar^2}\nnb \\[3mm] x \ge L\ \ :   & \dst\phi_n(x) = & \pm
d_n\,e^{-\rho(x-L)} &  \\[3mm] 0\ \le x\ \le L\ \ :   & \dst\phi_n(x) = & d_n
[\cos kx + \frac{\rho}{k}\sin kx]  & \dst\quad k^2 =\frac{2mE}{\hbar^2} 
\earr\eeq with
$$d_n  =  \frac{k}{\rho}\sqrt{\frac{2}{L}}
\frac{1}{\sqrt{(1+2/(\rho L))(1+k^2/\rho^2)}}.$$ The positive integer $n$ labels
the (finite for a given value of $V_0$) family  of solutions of the
transcendental equation :
$$ \tan\,(kL) = \frac{2k\rho}{k^2 - \rho^2}$$
\noindent and the $\pm $ corresponds to the (opposite) parity of the stationary 
state $n$, and to the relation 
$$\cos(kL)  +\frac{\rho}{k}\sin(kL) = \pm 1\,.$$

\noindent When $V_0$ is large, one finds for the spectrum  ($ \rho\ \simeq \nf\
,\ v_0 = 
\sqrt{\frac{2mV_0 L^2}{\hbar^2}}\ \gg \ 1\ ,\ k$ fixed) :
\beq\label{f3}
 k_n L \simeq n\pi(1 - 2/v_0)\ ,\ \ E_n \simeq E_n^{\nf} (1 -  4/v_0)\,
\eeq where the $ E_n^{\nf}$ 's are the infinite well energy levels
 (\ref{a3}), and for the stationary states :
\beqa\label{f4}
\phi_n(x\ \le\ 0) & \sim &
\sqrt{\frac{2}{L}}\left(\frac{n\pi}{v_0}\right)
\exp-v_0|x/L|\ \sim 0\nnb \\ 
\phi_n(x\ \ge\ L) &\sim &
\pm\sqrt{\frac{2}{L}}\left(\frac{n\pi}{v_0}\right)
\exp-v_0(x/L-1)\ \sim 0 \\ 
\phi_n(0\le\ x\ \le\ L) & \sim & \sqrt{2/L}
\left[\sin n\pi\frac{x}{L} + \left(\frac{n\pi}{v_0}\right)[\cos n\pi\frac{x}{L}
- 
\frac{1}{n\pi}\sin n\pi\frac{x}{L}]\right]\,,\nnb
\eeqa

In that (fixed energy) infinite limit of the finite well, we see that the 
standard boundary conditions $\,\phi(0)=\phi(L)=0\,$ are recovered.
 One could have considered a non-symmetric potential well such that 
$\,V(x)=V_0\,$ for $\,x<0\,$ and $\,V(x)=V_1\,$ for $\,x>L\,$  with $\,V_0\neq
V_1.$ Taking the limits $\,V_0\to\nf\,$ and $\,V_1\to\nf\,$  independently,
leads to the same conclusions as for the symmetric case $V_0 = V_1$ considered
here.

This result is hardly a surprise since for fixed $\,V_0\,$ we impose from the 
beginning the \underline{ continuity of the wave function and its first derivative} at 
$\,x=0\,$ and $\,x=L.$ The wave function in the classically forbidden region 
($\,x<0\,$ and $\,x>L$) is exponentially decreasing and is damped off to zero 
in the $\,V_0\to\nf\,$ limit. Combined with the continuity  of $\,\phi_n(x)\,$ 
at the points $\,x=0\,$ and $\,x=L\,$ this leads to $\,\phi(0)=\phi(L)=0\,$ (notice 
that in that limit the continuity of the first derivative of the wave  function
is lost).

In many textbooks \cite[vol. 1, p. 78]{CDL}, \cite[exercise 6.7, p. 396]{JMLL}, this limiting process is argued to select the  ``right"
boundary conditions for the self-adjoint extension of the  Hamiltonian. 
In the same spirit, it would be  tempting to consider the semi-axis case as a limit of  a step potential.  This selects uniquely the self-adjoint extension of the Hamiltonian such that $\phi(0)=0$ (Subsection 6.2). 
However, for any finite height, the momentum $P_x$ has a unique self-adjoint extension, while for an infinite height, 
$P_x$ has no self-adjoint extension at all (see Subsection 5.2)!

This discussion shows that an infinite potential cannot be simply described by the limit of a finite one.

\section {Concluding remarks} The aim of this article was twofold : first to
popularize the theory of  self-adjoint extensions of operators among people
learning and (or) teaching  quantum mechanics and second to point out 
some physical consequences which  could be checked by experiment. 

For example the new spectra for a particle in a box should lead to 
different low temperature behaviours of the specific heat, following 
the lines of \cite{Ro}, \cite{GA}. Similarly, the boundary effects 
computed in \cite{Be} should be examined anew.

Certainly the
examples considered here are too  simple, and are of questionable practical
feasability. Our hope is that  people will extend our analysis to the
differential operators acting in  three dimensional space which could lead to
more realistic physical  situations and put to light new phenomena : these
developements could  initiate the ``physics of self-adjoint extensions".

Moreover, as previously seen, an infinite potential cannot be simply described by the limit of a finite one. This enforces interest in the large class of  self-adjoint extensions described in this work : they  deserve further study since they are all on an equal footing with respect to the principles of quantum mechanics.

We have also emphasized in the previous Section the role of the symmetry properties  (resp. reality properties) of the boundary conditions when the potential has some symmetry properties  (resp. reality properties). Moreover, in subsection (5.3) we show that, in presence of an infinite discontinuity of the potential, the continuity of the wave function does not result from the principles of quantum mechanics.

Last, but not least, let us mention other difficult problems 
which are not thoroughly dealt with in the standard 
teaching of quantum mechanics : the definition of higher  powers
of operators ( to say nothing of their exponential !) and their 
commutators. This item  was
encountered in Section 2, where it was observed that $\,H^2\,$ is not  the
square of the operator $\,H\,.$ On the contrary, in subsection 7.3, we  have
exhibited a specific extension of $\,P^2\,$ which is  really the square of the
extension $\,P_{\tht}\,$ of $\,P.$

\appendix\newpage

\section{ Self-adjoint extensions of the momentum operator} Let us consider the
Hilbert space ${\cal H}=L^2(a,b)\,.$ The maximal domain  on which the operator 
$P=-i\hbar D$ has a well defined action has been called in Section.5
${\cal D}_{\rm max}(a,b)$. It is the linear space of functions $\psi(x)$ 
constrained by :
\brm
\item $\psi(x)$ is absolutely continuous \cite{absolutecontinuous} on $[a,b]$.
\item $\psi(x)$ and $\psi'(x)$ belong to $L^2(a,b)$.
\erm It is useful to introduce the quantity
\beq\label{identite} B(\psi,\phi)\equiv
\frac{1}{2i}\left[(P\psi,\phi)-(\psi,P\phi)\right]=\frac{\hbar}{2}[\ol{\psi}(b)\phi(b)
-\ol{\psi}(a)\phi(a)].\eeq

\subsection{The operator $\,P\,$ on the whole real axis} The Hilbert space is
${\cal H}=L^2({\mb R})$ and the maximal domain of P is 
${\cal D}_{\rm max}({\mb R})$.

One can prove that for any $\psi$ in  this maximal domain, one has :
$$\lim_{x\to\pm\nf}\psi(x)=0.$$ 
Note that this statement would not be  true under the single hypothesis $\psi\in
L^2({\mb R})$. The symmetry of $P$ is then, for
$\phi,\psi\in{\cal D}_{\rm max}({\mb R}),$ an  obvious consequence of
(\ref{identite}).  To prove that $\,(P,{\cal D}_{\rm max}({\mb R}))\,$ is indeed
self-adjoint, one  should show that, if $\,\phi\in L^2({\mb R})\,$ is such that
$$\forall\psi\in{\cal D}_{\rm max}({\mb R}),\qq\left|\int_{-\nf}^{+\nf}\,
\psi'(x)\ol{\phi}(x)\, dx\right|\leq  C\left(\int_{-\nf}^{+\nf}\,|\psi|^2\,
dx\right)^{1/2},$$ then $\,\phi\,$ belongs to $\,{\cal D}_{\rm max}({\mb R}).$
But it is easier  to check this using von Neumann's theorem, which was done in
Subsection 5.1. We have proven that the deficiency  indices are
$(0,0)$ and concluded that the operator $\,(P,{\cal D}_{\rm max}({\mb R}))\,$ is 
the {\sl unique} self-adjoint extension of $D$.

\subsection{The operator $\,P\,$ on the positive semi-axis} The Hilbert space is
${\cal H}=L^2(0,+\nf)$ and we take as domain
\beq\label{sa} {\cal D}_0(P)=\{\psi\in{\cal D}_{\rm max}(0,+\nf)\quad\mbox{and}
\quad \psi(0)=0\}.\eeq As in the previous subsection one can prove that
$\dst\lim_{x\to +\nf}\psi(x)=0.$
 Then the symmetry of the operator $\,P\,$ on $\,{\cal D}_0(P)\,$ follows  again
from relation (\ref{identite}).

The adjoint of $\,(P,{\cal D}_0(P))\,$ is given by
$$(P^{\dg}=P,\quad{\cal D}(P^{\dg})={\cal D}_{\rm max}(0,+\nf)).$$ The double
adjoint is simply
$$P^{\dg\dg}=P,\qq{\cal D}(P^{\dg\dg})={\cal D}_0(P),$$ which shows that
$(P,{\cal D}_0(P))$ is closed.

However, as we checked in Subsection 5.2, the deficiency indices are
$(1,0)$ and therefore, by von Neumann's theorem, 
$(P,{\cal D}(P))$ has no self-adjoint extension.

\subsection{The operator $\,P\,$ on a finite interval} The Hilbert space is now
${\cal H}=L^2(0,L)$ and we take
$$P=-i\hbar D,\qq {\cal D}_0(P)=\{\psi\in{\cal D}_{\rm max}(0,L),\quad 
\psi(0)=\psi(L)=0\}.$$ The symmetry of $\,P\,$ on $\,{\cal D}_0(P)\,$ follows
again from relation  (\ref{identite}). Its adjoint is
$$(P^{\dg}=P,\quad{\cal D}(P^{\dg})={\cal D}_{\rm max}(0,L)).$$ Let us notice
that the adjoint of $(P,{\cal D}(P^{\dg}))$ is 
$(P,{\cal D}_0(P))$ which implies its closedness. 

In Subsection 5.3, we have obtained the deficiency indices  
$(1,1)$ and,
from von Neumann's theorem, we know that the self-adjoint extensions are 
parametrized by $U(1)$ i.e. a phase $e^{i\tht}\,.$

\subsection{ Remarks}
\brm
\item In all cases we observe that the adjoint $\,(P^{\dg},{\cal D}(P^{\dg}))\,$ 
has for domain $\,{\cal D}(P^{\dg})={\cal D}_{\rm max}\,$ which is the largest 
domain in ${\cal H}$ in which $-i\hbar D$ is defined. It follows that the 
actual computation of the deficiency indices is always an easy task.
\item Let us observe that for symmetric operators one has the hierarchy
$$(P,{\cal D}(P))\quad\subset\quad(P^{\dg},{\cal D}(P^{\dg}))$$ which means that
the adjoint is the ``biggest". When self-adjoint extensions 
$(P,{\cal D}_{\tht})$ do exist they must lie in the in-between, according  to
the scheme
$$(P,{\cal D}_0(P))\quad\subset\quad(P,{\cal D}_{\tht})
\quad\subset\quad(P^{\dg},{\cal D}(P^{\dg}))$$
\item For further use, let us point out the useful theorem,  proved in
\cite[vol. 2, p. 90]{Na}, stating that for a differential  operator of order
$\,n\,$ with deficiency indices $\,(n,n)\,$ all of its  self-adjoint extensions
have a discrete spectrum.
\erm

\section{ The spectra of the Hamiltonian in a box} Starting from the boundary
conditions (\ref{r1}) we now derive the  equations giving the eigenvalues for
all the extensions $H_U.$

Let us consider the positive spectrum, the zero and negative ones being obtained
in the same way and, as a matter of fact, obtained by substitutions as indicated in 
(\ref{r14}).

 Denoting by
$\dst\  E=\frac{s^2}{L^2},$ with
$\ s>0,$ the eigenvalues  of $H_U$, and its eigenfunctions by
\beq\label{r2}
\dst \phi(s,x)=A\, e^{isx/L}+B\, e^{-isx/L},\qq 
\Phi=\left(\barr{l} A \\ [2mm]B \earr\right),\eeq one can easily check the
relations
\beq\label{r3}
\dst \left(\barr{l} L\phi'(0)-i\phi(0) \\ [2mm]  L\phi'(L)+i\phi(L) \earr\right)=
i{\mc L}(s)\Phi,\qq
\left(\barr{l} L\phi'(0)+i\phi(0) \\ [2mm]  L\phi'(L)-i\phi(L) \earr\right)=
i{\mc M}(s)\Phi,\eeq with the matrices
\beq\label{r4}
\dst {\mc L}(s)=\left(\barr{cc}
\dst s-1 & -s-1 \\ [2mm] \dst (s+1)e^{is} & -(s-1) e^{-is}\earr\right),\qq
{\mathcal M}(s)=\left(\barr{cc}
\dst s+1 & -s+1 \\ [2mm] \dst (s-1)e^{is} & -(s+1) e^{-is}\earr\right).\eeq The
determinants of these matrices are given by
$$\det\,{\mc M}(s)=2[i(s^2+1)\sin s-2s\cos s],\qq\qq \det\,{\mc
L}(s)=-\overline{\det\,{\mc M}(s)},$$ from which it follows that ${\mc L}(s)$
and ${\mc M}(s)$ have  vanishing determinant if and only if $s=0.$

Using these notations the equations for the eigenfunctions become
\beq\label{r5}
\dst\left({\mc L}(s)-U{\mc M}(s)\right)\Phi=0,   \eeq and for the spectra
\beq\label{r6}
\det\left({\mc L}(s)-U{\mc M}(s)\right)=0.   \eeq

To get a more explicit form of the eigenvalue equation let us use some  simple
relations valid for arbitrary $2\times 2$ matrices
\beq\label{r7} 2\det A=(\tr\, A)^2-\tr(A^2)\quad\Rightarrow\quad
\det(A-B)=\det A +\det B+\tr(AB)-\tr A\cdot\tr B.\eeq

For $s\neq 0$ we can write relation (\ref{r6}) as
\beq\label{r8}
\det({\mc L}(s){\mc M}^{-1}(s)-U)=0,\eeq where the matrix ${\mc L}{\mc M}^{-1}$
has the simple form
$$\dst {\mc L}(s){\mc M}^{-1}(s)=\frac 2{\det{\mc M}(s)}
\left(\barr{cc}\dst i(s^2-1)\sin s & -2s \\
 [3mm]-2s & i(s^2-1)\sin s\earr\right).$$ Subsequent use of (\ref{r7}) in
relation (\ref{r8}) and simple computations  lead to
\beq\label{r9}
\dst 2s\left[\cos s(1-\det U)-\,{\rm tr}\,(U\tau_1)\right]+ i\sin
s[(s^2+1)(1+\det U)-(s^2-1)\,{\rm tr}\,U]=0,\eeq valid for the positive non-zero spectrum.

The parametrization of the matrix $U$ given by 
(\ref{r10}), (\ref{r12})  simplifies relation (\ref{r9}) to
\beq\label{sp}\dst  2s\left[\sin\psi\cos s+\frac 1{2i}{\rm tr}\,(M\tau_1)\right]=
\sin s\left[(s^2+1)\cos\psi-\frac 12(s^2-1){\rm tr}\,(M)\right],\eeq a writing
which exhibits the reality of the eigenvalue equation. It also  displays a nice
invariance under the transformation
\beq\label{r11-bis} M\quad\to\quad  M'=e^{-\tht\tau_1/2i}\,M\,e^{+\tht\tau_1/2i},
\qq \tht\in[0,2\pi],\eeq as it leaves $\,{\rm tr}\,M\,$ and $\,{\rm
tr}(M\tau_1)\,$ unchanged. Let us  point out that this invariance is specific of
the spectra, not of  the eigenfunctions.

The strictly positive spectrum is then given by
\beq\label{spos} 2s\left[\sin\psi\cos s-m_1\right]=\sin
s\left[\cos\psi(s^2+1)-m_0(s^2-1)\right],
\qq E=\frac{s^2}{L^2}.\eeq The invariance (\ref{r11-bis}) explains why the
spectrum does not depend either of 
$\,m_2\,$ or of $\,m_3.$

An explicit solution of the eigenvalue equation
(\ref{spos}) is clearly  hopeless for the most general unitary matrix $U$.
Nevertheless there are  many special cases for which this can be achieved
explicitly. We therefore  classify the spectra as :
\brm
\item ``Simple" if the eigenvalue equation can be solved explicitly. This 
happens for two families :
$$ m_1=\sin\psi=0,\qq\qq{\rm or}\qq\qq  m_0=\cos\psi=0.$$
\item ``Generic" if this is not the case. Typically the ``generic" spectra  are
solutions of at least one transcendental equation and only their large 
$n$ behaviour can be obtained explicitly.
\erm

\subsection{First family of ``simple" spectra}  This first family corresponds
to
$\,\psi=0\,$ and $\,m_1=0\,$ and its  matrix $\,U\,$ has the form
\beq\label{ssp1}U=\left(\barr{cc} m_0-im_3 & -m_2 \\ [3mm] m_2 &
m_0+im_3\earr\right)\quad{\rm with}\quad m_0^2+m_2^2+m_3^2=1
\quad\Leftrightarrow\quad m\in S^2.\eeq The eigenvalue equation reduces to
$$\sin s\left[(1-m_0)s^2+1+m_0\right]=0.$$ Since $\,m_0\in[-1,+1]\,$  the factor
in front of the sine never vanishes,  so we get for spectrum
$$s_n=n\pi,\qq\qq n=1,2,\ldots$$

Note that the zero spectrum is easily checked to appear only for the
extension  with
$\,U=-{\mb I}\,,$ while the strictly negative spectrum is given by
$$\sinh r\left[(m_0-1)r^2+m_0+1\right]=0,$$ which has always a solution,
except for $\,m_0=\pm 1.$ We conclude to  the negative energy
$$\dst r^2=\frac{1+m_0}{1-m_0},\quad\longrightarrow\quad  E=-\frac
1{L^2}\,\frac{1+m_0}{1-m_0},\qq\qq m_0\in]-1,+1[.$$

\noindent {\bf Remark :} In this family two and only two extensions (with
$\,m_0=\pm 1$) are therefore  distinguished by the absence of negative energies
in their spectra. The  first one is
$$\dst \left\{\barr{c} U={\mb I} \\[3mm] \phi(0)=\phi(L)=0\earr\right.
\qq\longrightarrow\qq\left\{\barr{l}\ \  s_n=n\pi,\ n=1,2,\ldots,\\[3mm]
\ \ \dst\phi_n(x)=\sqrt{\frac 2L}\sin\left(n\pi\frac xL\right).\earr\right.$$
This is the ``standard" self-adjoint extension considered in the textbooks  on
quantum mechanics \cite[p. 109]{Gr},\cite[p. 300]{JMLL}.

The second one is
\beq\label{ssp2}
\dst \left\{\barr{c}U=-{\mb I}\ \\[3mm]\phi'(0)=\phi'(L)=0\earr\right.
\qq\longrightarrow\qq\left\{\barr{l}\ \ s_n=n\pi,\ n=0,1,\ldots,\\[3mm]
\ \ \dst\phi_n(x)=\sqrt{\frac 2L}\cos\left(n\pi\frac xL\right)\earr\right..\eeq

A different understanding of the absence of negative energies for these two 
extensions is given, using von Neumann theorem, in Subsection 7.3.

\subsection{Second family of ``simple" spectra.} This second family
corresponds to $\cos\psi=0,$ or equivalently 
$\,\psi=\pi/2,$ and $\, m_0=0.$ The corresponding matrix $\,U\,$ is
\beq\label{ssp3} U=\left(\barr{cc} m_3 & m_1-im_2 \\[3mm] m_1+im_2 &
-m_3\earr\right),\qq{\rm with}\qq m_1^2+m_2^2+m_3^2=1.\eeq Relation (\ref{spos})
reduces to
$$\cos s=m_1.$$ From (\ref{ssp3}) we know that $\,m_1\in[-1,+1].$ Excluding the values 
$\,m_1=\pm 1,$ discussed in the final remark, the positive spectrum is
$$s_n=\left\{\barr{l}+\cos^{-1}\,(m_1)+2n\pi,\qq n=0,1,\ldots \\[3mm]
-\cos^{-1}\,(m_1)+2n\pi,\qq n=1,2,\ldots\earr\right.\qq \cos^{-1}\,(1)=\pi/2.$$ As
already observed, these eigenvalues are independent of 
$\,m_2\,$ and $\, m_3,$ but this degeneracy affects only the spectra, not  the eigenfunctions.

Let us observe that for the particular case
$$U=\left(\barr{cc} 0 & \dst e^{-i\tht} \\[3mm] \dst e^{i\tht} & 0\earr
\right),\qq\qq
\tht\in]0,2\pi[$$ we have the full spectrum and eigenfunctions
$$\dst s_n=\tht+2n\pi,\qq\phi_n(x)=\frac 1{L}e^{2\pi i(n+\tht/2\pi)x/L},\qq 
n=1,2,\ldots$$
$$\dst s_n=-\tht+2n\pi,\qq\phi_n(x)=\frac 1{L}e^{2\pi i(-n+\tht/2\pi)x/L},\qq
n=0,1,\ldots$$ The exceptional cases $\,\tht=0\,$ and $\,\tht=\pi\,$ are
discussed in the next  remark. The important point is that these eigenfunctions
of $\,P^2\,$ are the  same as for $\,(P,{\cal D}_{\tht})\,$ given in Section 5.4

Note that the zero spectrum is easily checked to appear only for the
extension  with
$\,U=\tau_1,$ while the strictly negative spectrum given by 
$$\cosh r=m_1$$ 
is absent because from (\ref{ssp3}) we know that $\,m_1\in[-1,+1].$

\vspace{3mm}
\noindent {\bf Remark :} two extensions are distinguished by their doubly 
degenerate spectra. The first one corresponds to the periodic boundary 
conditions (the degeneracy of the energy $\,s_n\,$ is  denoted by
$\,g_n.$)
$$\left\{\barr{c} U=\tau_1\\[3mm]\phi(0)=\phi(L),\ 
\phi'(0)=\phi'(L)\earr\right.\quad\longrightarrow\quad
\left\{\barr{l}s_n=2n\pi,\quad n=1,2,\ldots\quad g_n=2,\\[3mm] s_0=0,\qq
g_0=1,\earr\right.$$ and the second one to the antiperiodic boundary conditions
$$\left\{\barr{c} U=-\tau_1\\[3mm]\phi(0)=-\phi(L),\  
\phi'(0)=-\phi'(L)\earr\right.\quad\longrightarrow\quad s_n=(2n+1)\pi,\quad
n=0,1,2,\ldots\quad g_n=2.$$ The periodic boundary conditions may have a
physical interpretation for  rotational degrees of freedom of molecules
\cite[vol. 2, p. 1202]{CDL}.

\subsection{The ``generic" spectra} We now exclude from our analysis the
extensions with ``simple" spectra.  Switching to the variable $\dst t=\tan\,\frac
s2,$ the eigenvalue equation  (\ref{spos}) becomes
\beq\label{sg1}
\dst\frac s{t^2+1}\left\{(m_1+\sin\psi)t^2+\frac
ts[\cos\psi(s^2+1)-m_0(s^2-1)]+m_1-\sin\psi\right\}=0.\eeq The overall factor
$\dst\,\frac 1{1+t^2}\,$ should not  be overlooked since it may vanish for
$\,t=\nf.$

We organise the discussion of the ``generic" spectra by  distinguishing three
different cases :
$$\hspace{-8cm}\bullet \ m_1=-\sin\psi\neq 0.$$ In this case the spectrum is
\beq\label{sg2}
\left\{\barr{l}
\dst \cot\,\frac s2=0\quad\longrightarrow\quad s=(2n+1)\pi,\qq 
n=0,1,2,\ldots,
\\ [3mm]
\dst\cot\,\frac s2=-\frac{(m_0-\cos\psi)s^2-(m_0+\cos\psi)}
{2s\,\sin\psi},\qq s>0.\earr\right.\eeq Notice that $\sin\psi$ cannot vanish
(because then $\,m_1=0\,$ and we are  back to the first family of ``simple"
spectra). The numerator vanishes  identically only for the second family of
``simple" spectra, so we  conclude that equation (\ref{sg2}) gives only
``generic" spectra.

$$\hspace{-8cm}\bullet \ m_1=\sin\psi\neq 0,$$ in which case we have
\beq\label{sg3}
\left\{\barr{l}
\dst\tan\,\frac s2=0\quad\longrightarrow\quad s=2n\pi,\qq n=1,2,\ldots,\\ [3mm]
\dst\tan\,\frac s2=\frac{(m_0-\cos\psi)s^2-(m_0+
\cos\psi)}{2s\,\sin\psi},\qq s>0.\earr\right.\eeq By the same argument as before
neither the numerator nor the denominator  can vanish, therefore equation
(\ref{sg3}) does give ``generic" spectra.

$$\hspace{-8cm}\bullet \ m_1\pm\sin\psi\neq 0,$$ in which case the discriminant
of equation (\ref{sg1}) can be written
$$\dst\Delta(s)=\left[(m_0-\cos\psi)s^2-(m_0+\cos\psi)\right]^2+
4s^2(m_2^2+m_3^2),$$ and is strictly positive because of the first term squared
(otherwise we are  back to the second family of ``simple" spectra).

The roots of
\beq\label{sg4}
\dst \tan\,\frac s2=\frac 1{2s(m_1+\sin\psi)}
\left\{(m_0-\cos\psi)s^2-(m_0+\cos\psi)\pm\sqrt{\Delta(s)}\right\},\eeq give
spectra which are certainly ``generic".

The equations giving the zero and the strictly negative spectrum can  also be
deduced as already explained in Section 6.

Let us conclude with a simple choice for the eigenfunctions
 
\beq\label{r16}
\barr{l}\dst A(s)=\alf\,(s-1)+[\ga\,e^{-is}-1](s+1),  \\ [3mm]
\dst B(s)=\alf\,(s+1)+[\ga\,e^{is}-1](s-1)=-A(-s),\earr 
\qq\qq U=\left(\barr{cc} \alf & \ga\\[3mm] \be & \de\earr\right).
\eeq This gives the relations
\beq\label{r17}
\dst\phi(s;x)=A(s)\, e^{isx/L}-A(-s)\, e^{-isx/L},\qq\phi(-s;x)=-\phi(s;x).\eeq
The simultaneous vanishing of $A$ and $B$ signals a doubly degenerate spectrum.

\section{Extensions preserving parity} 

The eigenfunctions (\ref{r17}) write in $u$ variable :
$$\tilde{\phi}_E(u)  = \phi_E(x) =A(s)e^{is/2}\, e^{isu}+B(s)e^{-is/2}\,
e^{-isu}.$$ Imposing the constraint (\ref{p1}) gives
\beq\label{p2} {\rm Im}\,(A(s)\ol{B}(s)e^{is})=0.\eeq It is important to observe
that this relation should hold only  when we take for $s$ the actual spectrum
given by relation (\ref{spos}).

Using for $\,A(s)\,$ and $\,B(s)\,$ the expressions given by (\ref{r16}),  and
after some algebra, one reduces the constraint (\ref{p2}) to
\beq\label{p3}
\barr{l}\hspace{-1cm} 2s\left[(\sin\psi\, m_0-\cos\psi\,m_3)\cos
s-m_0m_1-m_2m_3\right]=\\[5mm]
\hspace{+3cm}\sin s\left[(\cos\psi\, m_0+\sin\psi\,m_3)(s^2+1)-
(m_0^2+m_3^2)(s^2-1)\right].\earr\eeq The $\,m_0\,$ dependent terms disappear,
thanks to relation (\ref{spos}),  and we are left with
\beq\label{p4} m_3\,\sin\psi\left\{2s\left[\cos\psi\,\cos s+m_2\right]+
\sin s\left[\sin\psi(s^2+1)-m_3(s^2-1)\right]\right\}=0.
\eeq One can check, by enumeration of all 
the cases, that the coefficient between braces never vanishes for 
$\,m_3\neq 0.$

We conclude that all the parity preserving extensions are given by 
$\,m_3=0\,.$ \hfill{Q.E.D}

\bibliographystyle{plain}

\begin{thebibliography}{111}
\bibitem{expansion} Notice that the positive function (\ref{a4}) is nearly equal
to  the eigenfunction $\Psi_1$ as $b_1 = 0.99..,\ b_2 = -b_1/27\ ,\  b_3 =
b_1/125
\ldots$
\bibitem{absolutecontinuous} To make things simple we say that a function is
absolutely  continuous for $\,x\in[b,c]\,$ if it can be written in the form 
$\phi(x)=\int_a^x\,\psi(u)\, du,$ where $\psi(x)$ is absolutely integrable  for
any $\,x\in[b,c].$ Absolute continuity in a finite interval implies  uniform
continuity, whereas the converse is not true. The interested reader  is referred
to \cite[p.337]{KF}.
\end{thebibliography}

\end{document}